\documentclass[english,prl,twocolumn,superscriptaddress,nofootinbib,preprintnumbers]{revtex4}
\usepackage[T1]{fontenc}
\usepackage[latin1]{inputenc}
\usepackage{textcomp}
\usepackage{amssymb}

\makeatletter

\makeatletter

\newcommand{\be}{\begin{equation}}
\newcommand{\ee}{\end{equation}}
\newcommand{\bea}{\begin{eqnarray}}
\newcommand{\eea}{\end{eqnarray}}
\providecommand{\Or}{\mathcal{O}}
\providecommand{\eref}[1]{(\ref{#1})}

\newcommand{\D}{\mathcal{D}}

\newcommand{\BD}{\omega_{\rm \scriptscriptstyle BD}}

\newcommand{\rem}[1]{ }

\makeatother

\usepackage{babel}
\makeatother

\begin{document}

\title{Mimicking general relativity in the solar system}

\author{L. Amendola}

\affiliation{INAF/Osservatorio Astronomico di Roma, Viale Frascati 33, 00040 Monteporzio
Catone (Roma), Italy}

\author{C. Charmousis}

\affiliation{LPT, Universit� Paris--Sud, B�timent 210, 91405 Orsay CEDEX,
France}

\author{S. C. Davis}

\affiliation{Service de Physique Th�orique, Orme des Merisiers, CEA/Saclay,
91191 Gif-sur-Yvette Cedex, France}

\begin{abstract}
In order for a modified gravity model to be a candidate for cosmological
dark energy it has to pass stringent local gravity experiments. We
find that a Brans-Dicke (BD) theory with well-defined second order
corrections that include the Gauss-Bonnet term possess this feature.
We construct the generic second order theory that gives, to linear
order, a BD metric solution for a point-like mass source. We find
that the Eddington parameter $\gamma$, heavily constrained by time
delay experiments, can be arbitrarily close to the GR value of 1,
with an arbitrary BD parameter $\BD$. We find the region where the
solution is stable to small timelike perturbations. 
\end{abstract}
\maketitle
%\noindent \emph{Introduction}: 
Brans-Dicke (BD) theory is a simple modification of general relativity
\citet{BD} (see also \citet{gd} for generalisations) as it is a
single massless scalar-tensor theory whose only parameter is the kinetic
coupling term $\BD$, \begin{equation}
S_{{\rm BD}}=\int\sqrt{-g}\,\left[\Phi\, R-\frac{\BD}{\Phi}(\nabla\Phi)^{2}\right]-16\pi\mathcal{L}_{{\rm matter}}\,.\label{BD}\end{equation}
 Its GR limit is obtained for $\BD\rightarrow\infty$. BD gravity
breaks the strong equivalence principle and yields at local scales
differing Eddington parameters $\beta$ and $\gamma$ to those of
GR, which are strictly equal to 1. In particular the parameter $\gamma$,
which measures how much spatial curvature is produced by unit rest
mass (see \citet{will}), is given by $\gamma=(1+\BD)/(2+\BD)$. It
is strongly constrained by time delay experiments, such as the one
conducted with the Cassini spacecraft, which recently gave $|\gamma-1|\lesssim10^{-5}$~\citet{cassini}
(for a recent review and alternative methods to measure $\gamma$
see~\citet{will}). This implies $\BD>40000$, therefore the scalar
sector is very weakly coupled.

On the other hand modification of general relativity is possible or
even needed in order to explain effects on cosmological and galactic
scales, or at scales just beyond the solar system (such as the Pioneer
anomaly~\citet{pioneer}). At cosmological scales, some $10^{15}$
times bigger than solar system scales, supernovae data \citet{super}
entertain the possibility that GR may be modified at large distances
\citet{dgp}. Modification of GR is also being envisaged at galactic
scales in order to explain deviations from standard Newtonian gravity
in galactic rotation curves, as in MOND or Bekenstein-Sanders theor~\citet{mond}.
Scalars have been quite naturally introduced in order to mediate gravity
modification or even as sources of cosmological dark energy \citet{scalar}.
These modifications are well into the classical, infrared sector of
gravity at very low energies, very far from the the Planck scale UV
sector, where quantum gravity becomes important. Even from the point
of view of UV modifications, string theory predicts a zoo of scalars
that, if massless, would give BD-type phenomenology in the solar system.
It is therefore quite fair to say that there is increasing tension
between gravitational constraints imposed experimentally for weak
gravity in the solar system and laboratory tests, as well as strong
gravity from binary pulsars (see for example \citet{Damour}), and
on the other hand theories of modified gravity or dark energy that
aim to explain unexpected outcomes of novel experimental data. The
aim of this letter is to show that well motivated second order corrections
to simple BD theory (with no potential) can mimic GR at the solar
system scale, in the sense of giving $\gamma=1$ independently of
the parameter $\BD$. This does not mean that the toy scalar-tensor
theory in question and GR would not be distinguishable, quite the
contrary, on cosmological scales their phenomenology would be totally
different and even at the solar system level one could detect some
effect, most probably by carrying out an experiment to measure $\beta$,
as we will discuss in the concluding remarks.

%\noindent \emph{The model}: 
Our starting point is the general (modulo field redefinitions) scalar-tensor
Lagrangian of second order in powers of the curvature tensor which
has the unique property of giving second order field equations, \begin{eqnarray}
\mathcal{L}=\sqrt{-g}[f_{1}R-f_{2}(\nabla\phi)^{2}+\xi_{1}\mathcal{L}_{GB}+\xi_{2}G^{\mu\nu}\nabla_{\!\mu}\phi\nabla_{\!\nu}\phi\nonumber \\
{}+\xi_{3}(\nabla\phi)^{2}\nabla^{2}\phi+\xi_{4}(\nabla\phi)^{4}-2V{}]-16\pi G_{0}\mathcal{L}_{\mathrm{mattter}}\,.\label{action}\end{eqnarray}
 The theory \eref{action} is parametrised by the potential $V$
and couplings $f_{i},\xi_{i}$ which are all functions of the scalar
field $\phi$. The Gauss-Bonnet term is $\mathcal{L}_{GB}=R_{\mu\nu\sigma\rho}R^{\mu\nu\sigma\rho}-4R_{\mu\nu}R^{\mu\nu}+R^{2}$
and is a topological invariant in 4 dimensions for GR but not for
scalar-tensor (\ref{action}). Theories of the above form~\eref{action}
have been proposed as a solution to the dark energy problem~\citet{GBDE}.
For the case of $f_{1}=1$, it has been shown that solar system data
can impose severe restrictions on the couplings $\xi_{i}$~\citet{GBss},
which allows the range of possible gravity modification to be narrowed
down. Throughout this article we work in units with $c=1$. We will
assume $\xi_{3}=\xi_{4}=0$ in the following as a compromise between
introducing new free coupling functions in the theory and generality
of the setup. We choose to keep the non-minimal interaction terms
between the graviton and the scalar $\xi_{1}$ and $\xi_{2}$ rather
than higher order scalar corrections. When searching for solutions
we will assume also a vanishing potential $V$ (as in BD) since invoking
a large mass is an obvious but totally \emph{ad hoc} way to evade
solar system constraints and would suppress observable effects at
all scales. Our work also contrasts with `chameleon' models~\citet{chameleon},
in which the scalar's gravity is suppressed at local (but not cosmological)
scales by a suitably chosen potential that yields a background dependent
mass in eq. (\ref{action}). The higher order corrections are chosen
so that the gravitational propagator does not pick up extra degrees
of freedom. Therefore the resulting field equations will be a-priori
ghost-free around the Minkowski vacuum, second order in the derivatives,
with well-defined Dirac distributional terms.

In order to derive the post-Newtonian equations, we now assume a static,
spherically symmetric metric in isotropic coordinates, \begin{equation}
ds^{2}=-(1-2U)dt^{2}+(1+2\Upsilon)\delta_{ij}dx^{i}dx^{j}+\Or(\epsilon^{3/2})\,,\label{metric}\end{equation}
 where $U$ is the Newtonian potential and $\Upsilon$ is the leading
post-Newtonian spatial contribution. They are both functions of the
radial co-ordinate $r$ only and are assumed to be of the order of
the smallness dimensionless parameter $\epsilon=Gm_{\odot}/r$ where
$m_{\odot}$ is the solar mass and $r$ is a characteristic length
scale of the problem. For the solar system $\epsilon\lesssim10^{-5}$
for $r$ greater than the sun's radius. The post-Newtonian parameter
(PPN) we will be calculating is Eddington's parameter defined as $\gamma=\Upsilon/U$.
Matter energy density $\rho_{m}$ is given by the mass of the sun
and is as usual assumed to be a distributional source at $r=0$: $\rho_{m}=m_{\odot}\delta^{(3)}(x)$.
This is an excellent assumption given that the Schwarzschild radius
of the sun is of the order of 3 km compared to scales of the order
of astronomical units. Note however, that higher order terms in $\epsilon$
have to be included in (\ref{metric}) in order to calculate $\beta$
(for the advance of Mercury's perihelion for example). For the relativistic
experiment we will consider here, namely time delay, our expansion
is necessary and sufficient. Let us now define the operators, \begin{equation}
\Delta F=\sum_{i}F_{,ii}\,,\qquad\D(X,Y)=\sum_{i,j}X_{,ij}Y_{,ij}-\Delta X\Delta Y\,,\end{equation}
 which will be the technical tool essential for our analysis. For
functions with only $r$-dependence they reduce to $\Delta F=r^{-2}\partial_{r}(r^{2}\partial_{r}F)$
and $\D(X,Y)=-2r^{-2}\partial_{r}(r\partial_{r}X\partial_{r}Y)\,$
and in particular, \begin{eqnarray}
\D(r^{-n},r^{-m}) & = & \frac{2nm}{n+m+2}\Delta r^{-(n+m+2)}\,,\nonumber \\
\Delta r^{-n} & = & \frac{n(n-1)}{r^{n+2}}-\frac{4\pi n\delta^{(3)}(x)}{r^{n-1}}\,,\label{formula}\end{eqnarray}
 and thus we can easily evaluate the relevant distributional parts
associated with $\D$. We do not make any assumptions about the relative
sizes of $f_{i}$, $\xi_{i}$, $V$, or their derivatives (since we
expect the higher order terms to play a significant role), and instead
include the leading order in $\epsilon$ contribution from each term
in the field equations \begin{eqnarray}
f_{1}\Delta U & = & -4\pi G_{0}\rho_{m}+V+\frac{\Delta f_{1}}{2}-2\D(U+\Upsilon,\xi_{1})\nonumber \\
 &  & {}+\Or(\epsilon^{2},f_{2},\epsilon V,\epsilon\xi_{2},\epsilon^{2}\xi_{1})\label{Ueq}\\
f_{1}\Delta\Upsilon & = & -4\pi G_{0}\rho_{m}-\frac{V}{2}-\frac{\Delta f_{1}}{2}-2\D(\Upsilon,\xi_{1})+\frac{\D(\phi,\zeta_{2})}{4}\nonumber \\
 &  & {}+\Or(\epsilon^{2},\epsilon V,f_{2}\epsilon\xi_{2},\epsilon^{2}\xi_{1})\label{Yeq}\end{eqnarray}
 where we have defined $\xi_{2}=\partial_{\phi}\zeta_{2}$ and $f_{2}=\partial_{\phi}h_{2}$.
We will also expand $f_{1}$ to first order in $\epsilon$, $f_{1}=\Phi_{0}+\Or(\epsilon)$.
\rem{ %%%
The higher order terms we have dropped in the above equations are
$\Or(\epsilon^{2},\epsilon V,\epsilon\xi_{2},\epsilon^{2}\xi_{1})$.
} %%%
The scalar field equation on the other hand is globally of one order
higher and gives to leading order \begin{eqnarray}
 &  & \partial_{\phi}h_{2}\Delta\phi+\Delta h_{2}=2\partial_{\phi}V+2\partial_{\phi}f_{1}\Delta(2\Upsilon-U)\nonumber \\
 &  & {}-8\partial_{\phi}\xi_{1}\D(U,\Upsilon)+\D(U-\Upsilon,\zeta_{2})\nonumber \\
 &  & {}+\partial_{\phi}\zeta_{2}\D(U-\Upsilon,\phi)+\Or(\epsilon^{2},\epsilon V,f_{2}\epsilon,\epsilon^{3}\xi_{1},\epsilon^{2}\xi_{2})\,.\label{phieq}\end{eqnarray}
 \rem{ For $r$ only dependence \eref{phieq} becomes \begin{eqnarray}
\left(\phi'^{2}f_{2}r^{4}\right)' & = & 2r^{4}V'+2f'_{1}r^{2}\left(r^{2}(2\Upsilon-U)'\right)'+16\xi'_{1}r^{2}(rU'\Upsilon')'\nonumber \\
 & - & 2r^{2}(U-\Upsilon)'\left(r\phi'^{2}\xi_{2}\right)'-4\phi'^{2}\xi_{2}r^{5/2}\left(\sqrt{r}(U-\Upsilon)'\right)\end{eqnarray}
 } For comparison, BD theory ($V,\xi_{i}\equiv0$) has \begin{equation}
f_{1}=\Phi\equiv\Phi_{0}+\phi\,,\qquad f_{2}\equiv\frac{\BD(\Phi)}{\Phi}\approx\frac{\BD}{\Phi_{0}}+\Or(\phi)\label{bd}\end{equation}
 Since we are expanding the equations to the lowest non-trivial order,
we cannot estimate the second PPN parameter $\beta$, which requires
higher order terms in the metric coefficients. This loss in generality
in the metric coefficients is compensated by the great generality
of our solution.

It is useful to define $\eta\equiv1-\gamma$ whereupon the various
constants in the model are related by $\BD=-2+1/\eta$ and $G=2G_{0}/[(2-\eta)\Phi_{0}]$.
\rem{ \begin{eqnarray}
\gamma & = & \frac{1+\BD}{2+\BD}\,,\qquad\BD=-2+\frac{1}{\eta}\,,\\
G & = & \frac{(4+2\BD)G_{0}}{(3+2\BD)\Phi_{0}}=\frac{2G_{0}}{(2-\eta)\Phi_{0}}\,.\end{eqnarray}
 } We see that $\eta=0$ gives exactly GR. We are interested in gravitational
theories emanating from (\ref{action}), which while not identical
to general relativity, give almost identical predictions for the weak
field of the solar system. One approach to this problem would be to
solve the field equations of the previous section for a range of coupling
functions $\xi_{i}$, $f_{i}$, and then compare the resulting potentials
$U$ and $\Upsilon$, with those of Einstein gravity. We will not
take this approach since we have no interest in the solutions of the
field equations, except for the special cases where they give (to
this order in $\epsilon$) precisely the Newtonian result $U\approx\Upsilon\approx Gm_{\odot}/r$.
This in particular gives us agreement with tests of Newton's law from
planetary orbits. Therefore, instead of trying to find the metric
(\ref{metric}) which solves the field equations for given $\xi_{i}$,
$f_{i}$, we will inversely start by assuming the desired Newtonian
form of $U$ and $\Upsilon$, and then view~\eref{Ueq}--\eref{phieq}
as equations for the coupling functions $f_{i}$, $\xi_{i}$ parametrising
the theory (\ref{action}). As discussed above, we will now set $V=0$,
just as in the standard BD model and we allow the PPN parameter $\gamma=1-\eta$
to take any value. Hence we take \be U = \frac{Gm{\odot}}{r}
\, , \qquad{}\Upsilon = (1-\eta)\frac{Gm{\odot}}{r} \,
. \ee Note also that the effective gravitational coupling $G$ need
not be equal to the fundamental parameter for the gravitational coupling
of matter $G_{0}$. The general solution of \eref{Ueq}--\eref{phieq}
is then \begin{eqnarray}
f_{1}= & \Phi_{0} & +Gm_{\odot}\left(\frac{\Phi_{0}\eta+\lambda}{r}+\frac{2-\eta}{2}\int\frac{\partial_{r}S}{r}dr\right)\,,\nonumber \\
r^{4}(\partial_{r}\phi)^{2}f_{2} & = & (Gm_{\odot})^{2}[\Phi_{0}\eta+\lambda-2(\Phi_{0}+\lambda)\eta^{2}\nonumber \\
 &  & {}-3(1-4\eta+2\eta^{2})S-\eta(3-2\eta)r\partial_{r}S]\,,\nonumber \\
\xi_{1} & = & -\frac{\eta\lambda r^{2}}{8}+\int\frac{r^{2}\partial_{r}S}{16}dr\,,\nonumber \\
(\partial_{r}\phi)^{2}\xi_{2} & = & Gm_{\odot}[2\frac{\lambda}{r}(1-\eta)-\frac{3-2\eta}{2}\partial_{r}S\,]\,,\label{sol1}\end{eqnarray}
 where $\lambda$ is an arbitrary, dimensionful constant, obtained
by the distributional part appearing in the equations of motion (\ref{Ueq})--(\ref{phieq})
as the boundary condition at $r=0$. On the other hand, $S(r)$ is
an arbitrary function with the regularity condition $rS'=6S$ as $r\rightarrow0$,
i.e.\ $S=r\partial_{r}S=0$ at $r=0$. Viewing the above expressions
as the solution of ordinary inhomogeneous differential equations for
$f_{i}$, $\xi_{i}$, the function $S(r)$ parametrises the general
homogeneous solution of (\ref{Ueq})--(\ref{phieq}) whereas $\lambda$
parametrises the particular solution. The integrals range from $\infty$
to $r$. The gravitational coupling satisfies \begin{equation}
\frac{G_{0}}{G}=\Phi_{0}\left[1-\frac{\eta}{2}\right]-\frac{\lambda}{2}\left[1-4\eta+2\eta^{2}\right]\,.\label{Gsol}\end{equation}
 The above equations fully specify the couplings needed to reproduce
a PPN parameter $\gamma$ and an exactly Newtonian $1/r$ gravitational
potential. In fact $\lambda$ and $S$ now parametrise the theory
(\ref{action}). Setting $S=\lambda=0$ gives us pure BD (\ref{bd})
with $\phi=Gm_{\odot}\Phi_{0}\eta/r$. Setting on top of that $\eta=0$
gives GR. The key point however is that if we set $\eta=0$ keeping
$S$ and $\lambda$ non-zero we have \textit{the same post-Newtonian
limit as standard GR}, i.e.\ $\gamma=1$ in (\ref{metric}) without
the theory actually being GR. Indeed note that the corresponding kinetic
coupling $f_{2}$ can take arbitrary values parametrised by $\lambda$
and $S$. Indeed when the higher curvature terms are included, the
Newtonian potential is still proportional to $1/r$. For non-trivial
$S$, this is because all the corrections to standard gravity cancel
out, making the gravity modifications `invisible' to this order. On
the other hand, if $\lambda$ is non-zero, the corrections do not
cancel, but instead `mimic' Newtonian gravity. This can be seen from
the fact that the effective gravitational coupling $G$ receives a
$\lambda$ dependent correction~\eref{Gsol}. A similar effect was
found for $f(\mathcal{L}_{GB})$ gravity in~\citet{fG}, although
the resulting $\gamma$ was too large.

This fact is made clearer when we note that the above solution~\eref{sol1}
does not give a specific form for $\phi$. This is natural, since
by a change of variables, $\phi$ can be made to take any desired
form. Since we wish to express the functions in terms of $\phi$,
let us take \begin{equation}
\phi=\phi_{1}\frac{r_{g}}{r}\,.\end{equation}
 The constant $\phi_{1}$ simply corresponds to a re-scaling of $\phi$.
Defining $r_{g}=Gm_{\odot}$, the expressions~\eref{sol1} then
give \begin{eqnarray}
f_{1} & = & \Phi_{0}+\left(\Phi_{0}\eta+\lambda\right)\frac{\phi}{\phi_{1}}+\frac{2-\eta}{2\phi_{1}}\int\phi\partial_{\phi}S\, d\phi\,,\nonumber \\
f_{2} & = & \frac{1}{\phi_{1}^{2}}[\Phi_{0}\eta+\lambda-2(\Phi_{0}+\lambda)\eta^{2}\nonumber \\
 & - & 3(1-4\eta+2\eta^{2})S+\eta(3-2\eta)\phi\partial_{\phi}S]\,,\nonumber \\
\xi_{1} & = & \frac{r_{g}^{2}\phi_{1}^{2}}{8}\left[-\frac{\eta\lambda}{\phi^{2}}+\int\frac{\partial_{\phi}S}{2\phi^{2}}\, d\phi\right]\,,\nonumber \\
\xi_{2} & = & r_{g}^{2}\phi_{1}\left[\frac{2\lambda}{\phi^{3}}(1-\eta)+\frac{3-2\eta}{2\phi^{2}}\partial_{\phi}S\right]\,,\label{solp}\end{eqnarray}
 with the regularity conditions at $r=0$ now implying $S$ and $\phi\partial_{\phi}S$
tending to zero as $\phi\to\infty$. $S$ can then be expanded as
$S=\sum_{n\ge1}c_{n}\phi^{-n}$, obtaining the general asymptotic
solution to all orders in $\phi$. Note also that the higher order
couplings are $r_{g}$ dependent which follows from the fact that
they are of dimension length squared. In particular this means that
if we introduce the Gauss-Bonnet coupling constant $\alpha$ then
it is related via a multiplicative number to the only length scale
of the problem $r_{g}$, namely, $\alpha=-\phi_{1}^{2}\eta\lambda r_{g}^{2}/8$.
In fact the multiplicative constant is the hierarchy generated between
the classical scale $r_{g}$ and $\sqrt{\alpha}$.

To illustrate our result we consider the simplest case of $S=0$,
and take $|\eta|<10^{-5}$ to agree with solar system constraints.
Without loss of generality we set $\phi_{1}=\lambda+\Phi_{0}\eta$.
\rem{ We then have from (\ref{solp}) keeping leading order terms
in $\eta$, \begin{eqnarray}
f_{1} & = & \Phi_{0}+\phi\,,\qquad f_{2}=\frac{\BD}{\Phi_{0}}=(\Phi_{0}\eta+\lambda)^{-1}\,,\label{S0}\\
\xi_{1} & = & -\frac{\eta\lambda^{3}r_{g}^{2}}{8\phi^{2}}\,,\qquad\xi_{2}=\frac{2\lambda r_{g}^{2}[\lambda+\eta(\Phi_{0}-\lambda)]}{\phi^{3}}\,.\end{eqnarray}
 } We see that even with $\eta=0$ we have a BD Lagrangian (\ref{action})
with the additional term $\xi_{2}G^{\mu\nu}\nabla_{\mu}\phi\nabla_{\nu}\phi$
which can reproduce general relativity up to the first post-Newtonian
parameter $\gamma$. Therefore finding $\gamma=1$ does not guarantee
the absence of a scalar interaction even in this simplest of cases
since the scalar coupling $\BD$ is still freely given by $\Phi_{0}/\lambda$.
Note also that the strength of the scalar interaction $\BD$ is inversely
proportional to the strength of the higher order corrections, as parametrised
by $\lambda$.

%\noindent \emph{Stability}: 
We will now examine the stability of the solution (\ref{solp}) with
respect to time-dependent perturbations. We take $U\to U+\delta U$,
etc.\ and keep the leading order time derivatives of $\delta U$
(up to $\Or(\delta U)$, for linear gravity terms, and $\Or(\epsilon\delta U)$
for quadratic terms). For simplicity we will restrict ourselves to
the extreme case of solution~\eref{solp} with $\eta=0$ and $S\equiv0$
. The corresponding BD-like parameter for the higher order theory
is $\BD=\Phi_{0}/\lambda$. The perturbation equations then reduce
to\begin{eqnarray}
3\BD\ddot{\delta\Upsilon}-\frac{3}{2}\frac{\ddot{\delta\phi}}{\phi_{1}} & = & -\BD\Delta\,\delta U+\frac{1}{2}\Delta\frac{\delta\phi}{\phi_{1}}\label{stab1}\\
-\BD\Delta\,\delta\Upsilon+\frac{1}{2}\Delta\frac{\delta\phi}{\phi_{1}} & = & -\frac{1}{r^{2}}\partial_{r}\left(\frac{\partial_{r}(r^{3}\delta\phi)}{r\phi_{1}}\right)\\
-3\ddot{\delta\Upsilon}-\frac{\ddot{\delta\phi}}{\phi_{1}} & = & \Delta\delta U-\Delta\frac{\delta\phi}{\phi_{1}}+2r\partial_{r}\left(\frac{\partial_{r}(\delta U-\delta\Upsilon)}{r}\right)\label{stab3}\end{eqnarray}
 where we have assumed that the perturbations are more regular than
the leading order solution $1/r$ as $r\rightarrow0$. If \eref{solp}
is to be a viable gravity model, there needs to be a reasonable range
of parameters for which $\delta U$, etc.\ oscillate, rather than
growing over time. Substituting $\delta U(t,\overrightarrow{r})=\delta U(t)\exp i\overrightarrow{k}\overrightarrow{\cdot r}$
and analogously for $\delta\phi,\delta\Upsilon$, we find \begin{equation}
\ddot{\delta\phi}=-\frac{2\omega_{BD}+3}{2\omega_{BD}+9-\frac{18k^{3}r^{3}(kr-3)}{(k^{2}r^{2}+4kr-4)^{2}}}k^{2}\delta\phi\to\frac{2\omega_{BD}+3}{9-2\omega_{BD}}k^{2}\delta\phi\end{equation}
 (the limit is for $r\to\infty$). In the same limit, the time-dependence
of $\delta U,\delta\Upsilon$ is the same as for $\delta\phi$.

If $\BD>9/2$ or $\BD<-3/2$, the perturbations will oscillate for
large $r$, rather than grow exponentially, indicating that our gravitational
solution mimicking GR is classically stable for a reasonable range
of parameters, at least at first order. Although it might be that
including higher order terms in the perturbation expansion the oscillations
turn out to be of a growing nature, one has to be reminded that the
oscillations will be naturally damped by the emission of gravitational
waves and by the background cosmological expansion. A further constraint
comes from requiring positive gravitational coupling. For the above
case, \eref{Gsol} reduces to $G=(G_{0}/\Phi_{0})2\BD/(2\BD-1)$.
Hence $G>0$ implies $\BD<0$ or $\BD>1/2$, which are already covered
by the above ranges. We expect qualitatively similar results for more
general~\eref{solp} with $\eta$ or $S(\phi)$ non-zero, although
a proof of this is beyond the scope of this paper. Let us remember
that $S(r)$ is anyway an arbitrary function and one can always choose
it a posteriori in such a way to maintain stability.

%\noindent \emph{Discussion}.
In this paper we exhibited a sensible second order (in powers of the
curvature tensor not derivative) scalar-tensor theory which shares
some characteristics of ordinary BD or GR, in particular, well-defined
second order field equations, distributional boundary conditions and
well defined stable vacua. We found that such a theory, given the
right coupling functions, can mimic a GR Eddington parameter $\gamma$
exactly equal to $1$ with virtually no constraint on the kinetic
coupling $\BD$ (in ordinary BD theory actual measurements of $\gamma$
give $\BD>40000$). In this sense we saw that the inclusion of higher
order operators in the action can mimic GR with a scalar-tensor theory.
We do not view the solutions we have found (\ref{solp}) or even the
model in question (\ref{action}), as some fundamental scalar-tensor
theory; our aim was rather to see how robust were the solar system
predictions to higher order corrections. Our conclusion is that certain
solar system constraints known to rule out theories such as BD are
not as robust in their GR prediction as one might think. In fact similar
results have been shown for certain vector-tensor theories~\citet{jacob}
although one expects closer agreement with the Eddington parameters
in the case of vectors rather than scalars.

This does not mean that one cannot distinguish between such higher
order scalar tensor theories and GR. For a start the second PPN parameter
$\beta$ may not be unity for such theories, although if we allow
for the remaining higher order operators $\xi_{3},\xi_{4}$ in (\ref{action}),
in principle we have the mathematical flexibility in the equations
to again fix $\beta=1$ by solving for the coupling functions. The
main difference between GR, BD and these higher order theories is
that the coupling functions are dimensionful. Thus the relevant solutions
such as (\ref{solp}) will depend on the length scale of the solution,
namely $r_{g}$, times some dimensionless number whose magnitude will
determine the {}``fine-tuning'' one has to impose between the length
scales of the theory and the solar system. In other words we would
view the experimental error bars as hierarchies between the higher
order couplings and local scales where the experiment is carried out.
This relation may also be relaxed by allowing for the general second
order theory at the expense of introducing further free parameters
in the theory (\ref{action}). We further note that we have constructed
gravitational theories which exactly reproduce the Newtonian potential
for the sun: $U=r_{g}/r$. In fact it is perfectly acceptable to have
$U-r_{g}/r$ non-zero, but smaller than the experimental bounds from
planetary orbits. The above issues as well as a calculation of $\beta$,
other observational signatures of such higher order theories, in the
laboratory or in the solar system, and their cosmology are open interesting
questions which we hope shall be addressed in the near future.

\noindent \emph{Acknowledgments}: We thank G. Esposito-Farese for
detailed explanations on the exact solutions in BD theory.

\newcommand{\lcut}[1]{}


\begin{thebibliography}{10}
\bibitem{BD} C.~Brans and R.~H.~Dicke, %``Mach's principle and a relativistic theory of gravitation,''
 Phys.\ Rev.\ \textbf{124}, 925 (1961). %%CITATION = PHRVA,124,925;%%


\bibitem{gd} T.~Damour and G.~Esposito-Farese, %``Tensor multiscalar theories of gravitation,''
 Class.\ Quant.\ Grav.\ \textbf{9}, 2093 (1992). %%CITATION = CQGRD,9,2093;%%


\bibitem{will} C.~M.~Will, %{\em The confrontation between general relativity and experiment,}
 gr-qc/0510072; {\em Theory and experiment in gravitational physics},
\textit{Cambridge University Press}.

\bibitem{cassini} B.~Bertotti, L.~Iess and P.~Tortora, %{\em A test of general relativity using radio links with the Cassini
 %spacecraft,}
 Nature \textbf{425}, 374 (2003). %%CITATION = NATUA,425,374;%%


\bibitem{pioneer} J.~D.~Anderson, et al. %``Study of the anomalous acceleration of Pioneer 10 and 11,''
 Phys.\ Rev.\ D \textbf{65}, 082004 (2002)\lcut{ {[}gr-qc/0104064]};
%%CITATION = PHRVA,D65,082004;%%
 %``Indication, from Pioneer 10/11, Galileo, and Ulysses Data, of an Apparent
 %Anomalous, Weak, Long-Range Accelerattion,''
 Phys.\ Rev.\ Lett.\ \textbf{81}, 2858 (1998)\lcut{ {[}gr-qc/9808081]}.
%%CITATION = PRLTA,81,2858;%%


\bibitem{dgp}G.R. Dvali, G. Gabadadze and M. Porrati, Phys. Lett.
B \textbf{484}, 112 (2000); K. Koyama, R. Maartens, JCAP \textbf{0601},
016 (2006); P. Zhang et al. arXiv:0704.1932 {[}astro-ph];T. Koivisto
and F. Mota, Phys. Rev. D \textbf{73}, 083502 (2006); R. Maartens
and E. Majerotto, Phys. Rev. D \textbf{74}, 023004 (2006); A. Lue,
R. Scoccimarro and G.D. Starkmann, Phys. Rev. D \textbf{69}, 124015
(2004); B. Boisseau et al., Phys. Rev. Lett. \textbf{85} 2236 (2006);
L. Amendola, R. Gannouji, D. Polarski, S. Tsujikawa, Phys.Rev.D75,
083504 (2007). C.~Charmousis and A.~Papazoglou, JHEP \textbf{0807}
(2008) 062 {[}arXiv:0804.2121 {[}hep-th]]; C.~Charmousis, R.~Gregory
and A.~Padilla, JCAP \textbf{0710} (2007) 006 {[}arXiv:0706.0857
{[}hep-th]].

\bibitem{super} A.~G.~Riess \textit{et al.} {[}Supernova Search
Team Collaboration], %   {\em Observational Evidence from Supernovae for an Accelerating
%     Universe and a Cosmological Constant,}
 Astron.\ J.\ \textbf{116}, 1009 (1998)\lcut{ {[}astro-ph/9805201]};
%%CITATION = ASTRO-PH 9805201;%%
% {\em Type Ia Supernova Discoveries at $z>1$ From the Hubble Space Telescope:
%    Evidence for Past Deceleration and Constraints on Dark Energy Evolution,}
 Astrophys.\ J.\ \textbf{607}, 665 (2004)\lcut{ {[}astro-ph/0402512]};
%%CITATION = ASTRO-PH 0402512;%%
S.~Perlmutter \textit{et al.} {[}Supernova Cosmology Project Collaboration],
%  {\em Measurements of Omega and Lambda from 42 High-Redshift Supernovae,}
 Astrophys.\ J.\ \textbf{517}, 565 (1999)\lcut{ {[}astro-ph/9812133]}.
%%CITATION = ASTRO-PH 9812133;%%


\bibitem{mond} J.~D.~Bekenstein, %``Relativistic gravitation theory for the MOND paradigm,''
 Phys.\ Rev.\ D \textbf{70} (2004) 083509 {[}Erratum-ibid.\ D \textbf{71}
(2005) 069901]\lcut{ {[}astro-ph/0403694]}; %%CITATION = PHRVA,D70,083509;%%
 J.~D.~Bekenstein and R.~H.~Sanders, %``A Primer to Relativistic MOND Theory,''
 astro-ph/0509519. %%CITATION = ASTRO-PH/0509519;%%


\bibitem{scalar}C. Wetterich, A\&A, 301, 321 (1995); R. R. Caldwell,
R. Dave, P.J. Steinhardt, Phys. Rev. Lett., 80, 8, (1998).

\bibitem{Damour} T.~Damour and G.~Esposito-Farese, %``Tensor-scalar gravity and binary-pulsar experiments,''
 Phys.\ Rev.\ D \textbf{54} (1996) 1474\lcut{ {[}gr-qc/9602056]}.
%%CITATION = PHRVA,D54,1474;%%


\bibitem{GBDE} L.~Amendola, C.~Charmousis and S.~C.~Davis, %{\em Constraints on Gauss-Bonnet gravity in dark energy cosmologies,}
 JCAP \textbf{0612}, 020 (2006)\lcut{ {[}hep-th/0506137]}; %%CITATION = JCAPA,0612,020;%%
%\bibitem{mota}
 T.~Koivisto and D.~F.~Mota, %{\em Cosmology and Astrophysical Constraints of Gauss-Bonnet Dark Energy,}
 Phys.\ Lett.\ B \textbf{644}, 104 (2007)\lcut{ {[}astro-ph/0606078]};
%%CITATION = PHLTA,B644,104;%%
 %{\em Gauss-Bonnet quintessence: Background evolution, large scale
 %  structure and cosmological constraints,}
 Phys.\ Rev.\ D \textbf{75}, 023518 (2007)\lcut{ {[}hep-th/0609155]};
%%CITATION = PHRVA,D75,023518;%%
 B.~M.~Leith and I.~P.~Neupane, %{\em Gauss-Bonnet cosmologies: Crossing the phantom divide and the transition
 %from matter dominance to dark energy,}
 hep-th/0702002. %%CITATION = HEP-TH/0702002;%%


\bibitem{GBss} G.~Esposito-Farese, %{\em Scalar-tensor theories and cosmology and tests of a
 %quintessence-Gauss-Bonnet coupling,}
 gr-qc/0306018; %{\em Tests of scalar-tensor gravity,}
 AIP Conf.\ Proc.\ \textbf{736}, 35 (2004)\lcut{ {[}gr-qc/0409081]};
T.~P.~Sotiriou and E.~Barausse, %{\em Post-Newtonian expansion for Gauss-Bonnet gravity,}
 gr-qc/0612065; L.~Amendola, C.~Charmousis and S.~C.~Davis, %``Solar System Constraints on Gauss-Bonnet Mediated Dark Energy,''
 JCAP \textbf{0710}, 004 (2007)\lcut{ {[}0704.0175 {[}astro-ph]]}.
%%CITATION = JCAPA,0710,004;%%


\bibitem{chameleon} P.~Brax, C.~van de Bruck, A.~C.~Davis, J.~Khoury
and A.~Weltman, %{\em Detecting dark energy in orbit: The cosmological chameleon,}
 Phys.\ Rev.\ D \textbf{70}, 123518 (2004)\lcut{ {[}astro-ph/0408415]}
%%CITATION = PHRVA,D70,123518;%%


\bibitem{fG} S.~C.~Davis, %``Solar System Constraints on f(G) Dark Energy,''
 0709.4453 {[}hep-th]. %%CITATION = ARXIV:0709.4453;%%


\bibitem{jacob} B.~Z.~Foster and T.~Jacobson, %``Post-Newtonian parameters and constraints on Einstein-aether theory,''
 Phys.\ Rev.\ D \textbf{73} (2006) 064015\lcut{ {[}gr-qc/0509083]}.
%%CITATION = PHRVA,D73,064015;%%

\end{thebibliography}
\end{document}